 \documentclass[twocolumn,showpacs,preprintnumbers,amsmath,amssymb,pra]{revtex4}
\usepackage{graphicx}
\begin{document}
\title{Quantum Entanglement in the Two Impurity Kondo Model}
\author{Sam Young \surname{Cho}}
\email{sycho@physics.uq.edu.au}
\author{Ross H. \surname{McKenzie}}
\email{mckenzie@physics.uq.edu.au}

\affiliation{Department of Physics, The University of Queensland,
             4072, Australia}

\date{\today}

\begin{abstract}
 In order to quantify quantum entanglement in two impurity Kondo systems,
 we calculate the concurrence, negativity, and von Neumann entropy.
 The entanglement of the two Kondo impurities is shown to be determined
 by two competing many-body effects, the Kondo effect and 
 the Ruderman-Kittel-Kasuya-Yosida (RKKY) interaction, $I$.
 Due to the spin-rotational invariance of the ground state,
 the concurrence and negativity are uniquely determined by 
 the spin-spin correlation between the impurities.
 It is found that there exists a critical minimum value of 
 the antiferromagnetic correlation
 between the impurity spins
 which is necessary for entanglement of the two impurity spins. 
 The critical value
 is discussed in relation with the unstable fixed point
 in the two impurity Kondo problem.
 Specifically, at the fixed point there is no entanglement
 between the impurity spins.
 Entanglement will only be created 
 (and quantum information processing (QIP) be possible)
 if the RKKY interaction exchange energy, $I$,
 is at least several times larger than the Kondo temperature, $T_K$.
 Quantitative criteria for QIP
 are given in terms of the impurity spin-spin correlation. 
\end{abstract}
\maketitle


 \section{Introduction}


The potential of quantum information
processing and quantum communication has led to
numerous proposals of specific
material systems for the creation and
manipulation of entanglement in solid state qubits \cite{Loss,Kane}.
Condensed matter systems have several appealing features:
(i) natural qubits such as single spin-1/2,
(ii) the dream of scaleability found in the solid state technology
which is the basis of classical computers,
 and (iii) the presence of
strong interactions between qubits, such as spin exchange,
which can create entanglement.
Furthermore, even when there is
no direct interaction between qubits, the
interaction of the individual qubits with
their environment can lead to
an indirect interaction between qubits \cite{Fisher}.
A concrete example of such an indirect interaction is the 
Ruderman-Kittel-Kasuya-Yosida (RKKY)
 interaction \cite{RK} between two localized spins
interacting with the itinerant spins in a metal.
This has led to several recent proposals to
use the RKKY interaction to produce and manipulate
entanglement in solid state qubits \cite{Glazman,Marcus,Piermarocchi,Ardavan}.
                                                                                
In considering these proposals for solid state
quantum information processing it is important
to bear in mind some results from quantum information
theory concerning the entanglement in
mixed states.
Even though entangled states result from interactions
and exhibit certain correlations (e.g., antiferromagnetic
interactions can produce singlet states which exhibit
antiferromagnetic correlations) such interactions
and correlations are necessary but not sufficient
for the presence of entanglement.
 In particular, Werner \cite{Werner} defined a sub-class of mixed states
of pairs of qudits
that had two particularly interesting sub-families.
One family of states  had ``classical'' correlations in the sense that
the two-qudit density matrix could be written
as a convex combination of product (i.e., unentangled) states.
Such states can be modeled by a hidden-variable theory
and satisfy Bell's inequalities.
A second distinct subfamily were entangled
but could be modeled by a hidden variable theory.
In this paper, we consider the implications of this
for the specific case of the two-impurity Kondo model,
which describes the interaction of
two localized spin-1/2's (qubits) interact via the
Heisenberg exchange interaction with the itinerant
electrons in a metal. We investigate how the competition between
the Kondo effect \cite{Hewson} and the RKKY interaction determines the
parameter regime for which entanglement of the two qubits
can occur.
Although we focus on this specific system many
of the results and concepts considered can
be readily adapted
to other solid state qubit systems.
For example, this is another example of how the entanglement
in the whole system is ``shared'' \cite{Dawson}: 
the extent to which two qubits can
be entangled with each other is limited by how entangled the individual
qubits are with the environment.

 Manipulation of many-body quantum states in solid state physics
 has come to reality.
 For example,
 the Kondo effect \cite{Kouwen}
 and superconducting qubits \cite{superconductor}
 have been realized experimentally in a controllable manner.
 For a quantum dot (QD) fabricated
 in a semiconductor two-dimensional electron gas (2DEG) system,
 system parameters can be varied in a tunable manner \cite{Bird}
 to explore various many-body effects in previously inaccessible regimes.
 Electron transport through QD's in the unitary limit
 has manifested that
 the ground state is a many-body Kondo singlet \cite{Yosida66} 
 as a result of the Kondo resonance \cite{Wiel,Ji}.
 This means that the localized magnetic moment is
 entangled with the itinerant electrons. 
 Further,
 it has recently been proposed that a tunable  
 RKKY interaction
 could be used to entangle two spatially separated spins
 and perform quantum information processing (QIP) electrically 
 \cite{Glazman,Marcus}
 or optically \cite{Piermarocchi} 
 in coupled QDs  or with endohedral fullerenes 
 inside carbon nanotubes \cite{Ardavan}.
 Varying the RKKY interaction,
 to induce the effect of transitions between
 different ground states 
 has been theoretically investigated
 \cite{Simon,Vavilov}.
 Furthermore, quantifying entanglement in quantum many-body systems
 has recently been investigated \cite{Sorensen,Osterloh,Osborne,Ghosh,
 Hines03,Costi,Barnum,Falco,Hines04,Jordan,Wu,Vidal04,Hines05,Yang}. 
 Motivated by a recent experiment of nonlocal spin control 
 in a coupled-QD system \cite{Marcus},
 it is important to understand
 how two spatially separated spins are entangled by
 tunable quantum many-body effects.
 To help the answer, 
 we quantify 
 quantum entanglement in two impurity Kondo systems.
 The outline of the paper is as follows.
 A general expression for the reduced density matrix for 
 the two impurity spins is given in terms of the spin-spin correlation. 
 It is found that to be entangled 
 the two impurity spins need a minimum non-zero antiferromagnetic (AFM) 
 correlation determined by 
 the competition between the Kondo effect and the RKKY interaction. 
 We point out that
 at the unstable fixed point \cite{Jones88} in the two impurity Kondo problem
 the AFM correlation has this critical value.

 \begin{figure}
 \vspace*{4.00cm}
 \includegraphics{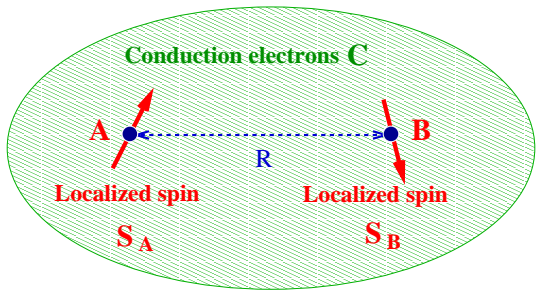}
 \caption{(Color online) Two impurity Kondo system.
          The total system may be regarded as a combined system containing 
          three subsystems, $A$, $B$, and $C$. 
          Two 
          localized spins $A$ and $B$ are separated by a distance $R$,
          and $C$ is the conduction electrons.
          These three subsystems interact with one another both directly 
          and indirectly.
          For example, the localized spins interact directly
          with the conduction electrons by the spin exchange interaction $J$ 
          and indirectly with each other by the RKKY interaction $I$
          which is mediated by 
          the conduction electrons. 
          }
 \label{2spin}
 \end{figure}

 \section{The two impurity Kondo model}

 The two impurity Kondo model describes 
 two localized spins interacting with 
 itinerant conduction electrons. 
 One may then suppose that the total system has three subsystems,
 consisting of the two localized spins ($A$ and $B$) 
 and the conduction electrons ($C$) as the environment (See Fig. \ref{2spin}).
 The Hamiltonian describing the two impurity Kondo model
 is 
 \begin{equation}
  H = H_C - J \Big( {\bf S}_A \cdot {\bf s}_c(A) 
         + {\bf S}_B \cdot {\bf s}_c(B) \Big),
 \label{model}
 \end{equation}
 where $H_C=\sum_{k\sigma} \varepsilon_k c^\dagger_{k\sigma}c_{k\sigma}$ 
 denotes the Hamiltonian for the conduction electrons.
 $J$ is the spin exchange coupling between the impurity spins
 ${\bf S}_{A(B)}$
 and the conduction-electron spin densities, 
 ${\bf s}_c({\bf R})=\frac{1}{2N_e}\sum_k c^\dagger_{k\sigma}
 \mbox{\boldmath $\sigma$}_{\sigma\sigma'}
 c_{k'\sigma'} \exp[i({\bf k}-{\bf k}')\cdot {\bf R}]$,
 at the impurity sites $A$ and $B$,
 where $N_e$ is the number of different ${\bf k}$ vectors.
 The relevant energy scale governing a single spin impurity model 
 is the Kondo temperature \cite{Hewson}, 
 $T_K \approx D \sqrt{J\varrho_F}\exp[-1/J\varrho_F]$,
 with the conduction band width, $D$, and 
 the single-particle density of states at the Fermi energy $\varrho_F$.
 The conduction electron spins
 mediate a spin exchange interaction
 between the two spatially separated impurity spins,
 the RKKY interaction, even though it does not explicitly appear 
 in Eq. (\ref{model}) of the Hamiltonian. 
 To second order in $J$, the RKKY interaction between
 two impurity spins can be described by the Hamiltonian 
 \begin{equation}  
   H_{RKKY} =  I(R) \; {\bf S}_A \cdot {\bf S}_B,
 \end{equation}  
 where $I(R)$ characterizes the effective spin-exchange interaction
 between the two impurity spins depending on the distance $R$.
 The exchange interaction varies
 as 
 $I(R)=4\pi J^2 \varepsilon_F F_{1,3}(2k_F R)$ with
 $F_3(x) = (\sin x - x\cos x)/x^4$ 
 in three dimensions \cite{Kittel}
 and $F_1(x) = -(1/4)\int^\infty_x dy \sin y/y$ in one dimension \cite{Yafet}.
 $\varepsilon_F$ and $k_F$ are the Fermi energy and the Fermi wavevector,
 respectively.
 Recently the RKKY interaction in single-walled nanotubes has been studied
 \cite{Shenoy} theoretically.
 Note that the sign of $I(R)$ depends on the distance $R$ between
 the two impurity spins.
 AFM coupling occurs for $I(R)>0$
 and ferromagnetic (FM) coupling occurs for $I(R)<0$.
 The competition between the RKKY interaction and Kondo effect
 determines the characteristics of the system 
 by the ratio of the relevant energy scales, $I(R)/T_K$.
 For instance, for a strong ferromagnetic RKKY interaction, $|I| \gg T_K$,
 the two stage Kondo effect \cite{Jayaprakash} is seen 
 in the temperature dependence of susceptibility,
 there are three distinct temperatures
 at which the susceptibility decreases. 
 The ratio $I(R)/T_K$ can be varied by changing $J$ or $R$.
 \section{Reduced density matrix for the two impurity spins}

 At zero temperature $(T=0)$, the total system
 should be in a ground state, $\left|\Psi_G\right\rangle$,
 which is pure.
 The ground state should be a spin singlet \cite{Jones88}.
 This means that it is invariant under joint rotation of all the spins.
 To quantify entanglement between the two localized spins, 
 let $\rho=\left|\Psi_G\right\rangle\left\langle\Psi_G\right|$ 
 be the density matrix for the ground state of the total system.
 Although the total system is in a pure state, 
 the two localized spins are in a mixed state.
 For any two qubits, here two Kondo impurity spins ${\bf S}_A$ and ${\bf S}_B$,
 the density matrix 
 can be written in the form \cite{Nielsen},
   \begin{equation}
    \rho_{AB} = \mathrm{Tr}_{C}(\rho)
      = \frac{1}{4}\sum_{\alpha,\beta=0,x,y,z}
         r_{\alpha\beta} \; \sigma^\alpha_A \otimes \sigma^\beta_B,
   \end{equation}
  where the coefficients in this operator expansion
  are determined by the relation
  \begin{equation}
   r_{\alpha\beta} = \mathrm{Tr}(\sigma^\alpha_A\; \sigma^\beta_B\; \rho_{AB})
       = \langle \sigma^\alpha_A \; \sigma^\beta_B \rangle.
  \end{equation}
 $\sigma^0_j$ and $\sigma^\mu_j$ $(\mu=x,y,z)$ are the identity matrix
 and the Pauli matrices, respectively and $j=A$ and $B$.
 The reduced (four by four) density matrix $\rho_{AB}$ 
 is obtained from $\rho$ by taking
 the partial trace over the states of subsystem $C$ (conduction electrons).
 All influences of the direct and indirect interactions 
 between the two Kondo spins
 are contained 
 in the correlation functions defined by the coefficients $r_{\alpha\beta}$.

 For the two Kondo spins,
 we derive a general expression for the reduced density matrix
 which is valid when any total system considered
 satisfies the following symmetries.
  The reflection symmetry of the system implies that $\rho_{AB} = \rho_{BA}$.
  Since 
  the Hamiltonian describing the system is real, $\rho^*_{AB} = \rho_{BA}$.
  Furthermore, 
  if the ground state is a total spin singlet, i.e.,
  spin rotationally invariant, then $r_{\alpha\beta}=0$ if $\alpha \neq \beta$
  and
  $r_{xx}=r_{yy}=r_{zz}=r$.
  The symmetries require that the only nonzero coefficients
  in the operator expansion are $r_{00}$, $r_{xx}$, $r_{yy}$, and $r_{zz}$.
  In addition, $r_{00}=1=\mathrm{Tr}(\rho_{AB})$
  because the density matrix must have trace unity. 
  The reduced density matrix may depend only on the distance $R$ between
  the two Kondo impurity spins ${\bf S}_A$ and ${\bf S}_B$
  because in our study the indirect RKKY interaction between 
  the Kondo impurity spins is mediated by the conduction electrons. 
  In the basis
       $\left\{ \left|\uparrow\uparrow \right\rangle, 
           \left|\uparrow\downarrow \right\rangle,
           \left|\downarrow\uparrow \right\rangle,
           \left|\downarrow\downarrow \right\rangle \right\}$,
  the reduced density matrix $\rho_{AB}$ can be rewritten 
  entirely in terms of the spin-spin correlation function, 
  $f_s \equiv \langle {\bf S}_A \cdot {\bf S}_B \rangle=3r/4$,
  as follows,
  \begin{eqnarray}
  \rho_{AB} 
   &=&
  \frac{1}{4}\Big( {\mathbf I} + r
  \sum_{\alpha=x,y,z} \sigma^{\alpha}_A \otimes \sigma^{\alpha}_B \Big),
 \label{R1}
 \end{eqnarray}
 where ${\mathbf I}$ denotes the four by four identity matrix.

 To get more insight into the entanglement for the 
 two impurity Kondo system,
 we rewrite the reduced density matrix
 in the Bell basis of  maximally entangled states
 $\left\{ \left|\Psi^\pm \right\rangle, 
 \left|\Phi^\pm \right\rangle \right\}$,
 where
  $ \left|\Psi^\pm\right\rangle 
   =  \frac{1}{\sqrt{2}}
  ( \left|\uparrow\downarrow \right\rangle \pm 
         \left|\downarrow\uparrow \right\rangle)$ and
  $ \left|\Phi^\pm\right\rangle 
   =  \frac{1}{\sqrt{2}}
  ( \left|\uparrow\uparrow \right\rangle 
      \pm \left|\downarrow\downarrow \right\rangle)$.
 Note that $\left| \Psi^-\right\rangle$ is the spin singlet state.
 The result is
 \begin{equation}
  \rho_{AB}
 = 
  p_s |\Psi^-\rangle \langle \Psi^-| \!
 +p_t \Big( |\Psi^+\rangle \langle \Psi^+| \!
            +\! \sum_{i=\pm} |\Phi^i\rangle \langle \Phi^i| \Big), 
 \label{R2}
 \end{equation}
 where
 $p_s = 1/4-f_s$ 
 and 
 $p_t = 1/4+f_s/3$.
 This state is a singlet-triplet mixture.
 The spin-spin correlation function of any two spins is bounded: 
 $ -3/4 \leq f_s \leq 1/4$.
 Thus,
 the probabilities for the singlet and three triplet states are
 $ 0 \leq p_s \leq 1 $ and $ 0 \leq p_t \leq 1/3$.
 The spin-spin correlation determines the properties of the state
 for the two spins.
 The probability of the two spins being in a singlet state
 is $P(S)=p_s=1/4-f_s$
 and the total probability of triplet states is $P(T)=3p_t=3/4+f_s$.
 In the limit of a pure AFM singlet of the two spins, 
 i.e., $f_s=-3/4$,
 the two spins are in a maximally entangled state;
 $\rho_S = \left| \Psi^- \right\rangle \left\langle \Psi^- \right|$,
 with $P(S)=1$ and $P(T)=0$.
 While
 in the limit of no singlets, 
 i.e., $f_s=1/4$,
 the two spins are in 
 a equal mixture of three triplet states;
  $\rho_{T} = 
  \frac{1}{3} \left( |\Psi^+\rangle \langle \Psi^+| 
                   +\sum_{i=\pm} |\Phi^\pm\rangle \langle \Phi^\pm|\right)$,
 with $P(S)=0$ and $P(T)=1$.
 However, the concurrence/negativity for this particular mixture of 
 triplet states is zero as will be discussed below.
 When the spin-spin correlation vanishes, i.e.,
 $f_s= 0$,
 the reduced density matrix becomes $\rho_{AB} = \frac{1}{4} {\bf I}$,
 the totally mixed density matrix
 which is ``garbage" for QIP.
 Then there is no entanglement between the two spins.
 In this case,
 the probabilities for the singlet and triplet states
 are 
 $P(S)=1/4$ and $P(T)=3/4$.
 When the probabilities for the singlet and triplet states
 are equal, i.e., $P(S)=P(T)=1/2$,
 the spin-spin correlation is $f^c_s=-1/4$.
 The state for the two localized spins 
 can be regarded as an equal mixture of the total spin
 of impurities $S_{\rm imp}=0$ and $S_{\rm imp}=1$.

 \section{Concurrence/Negativity and a critical value of correlation} 

 $\rho_{AB}$ is actually 
 a Werner state \cite{Werner} and can be written as 
 \begin{eqnarray}
  \rho_{AB} \!\!
  &=& \!
   \frac{4p_s-1}{3} |\Psi^-\rangle \langle \Psi^-| 
   +\frac{1-p_s}{3} {\bf I } .
 \label{R3}
 \end{eqnarray}
 This state is characterized by a single parameter $p_s$ called
 the fidelity because 
 $p_s=\langle\Psi^-|\rho_{AB}|\Psi^-\rangle$
 measures the overlap of the Werner state with
 the spin singlet Bell state.
 One measure of entanglement is the concurrence \cite{Wootters98}.
 For the Werner state $\rho_{AB}$,
 the concurrence is given by \cite{Wootters98}
 \begin{eqnarray}
 C(\rho_{AB}) = \mathrm{max}\left\{ 2p_s-1, \; 0 \right\}.
 \label{R4}
 \end{eqnarray}
 For $ 0 \leq p_s \leq 1/2$ (i.e., 
 $ -1/4 \leq f_s  \leq 1/4$), 
 the concurrence is zero and the reduced density matrix 
 can be written as a convex combination of  
 (disentangled) product states.
 For $ 1/2 \leq p_s \leq 1$ (i.e., 
 $ -3/4 \leq f_s \leq -1/4$), 
 the concurrence ranges from zero 
 to one (a maximally entangled state), 
 and it is related to 
 the spin-spin correlation function monotonically.
 Therefore, at $p^c_s=1/2$, 
 there exists a critical value of the spin-spin correlation,
 $f^c_s= -1/4$, 
 separating entangled states from unentangled states. 
 In a quantum spin system, 
 a critical value of spin-spin correlation has been discussed 
 for a system consisting of a spin $S$ and a spin 1/2 \cite{Schliemann}. 
 Another important measure of entanglement 
 is the negativity  $N(\rho_{AB})$ \cite{Zyc,Vidal}.
 Similarly to the concurrence, the negativity ranges from zero to one. 
 The negativity of the Werner state \cite{Lee} is equal to 
 the concurrence, 
 \begin{equation}
  N(\rho_{AB})=C(\rho_{AB}).
 \label{R5}
 \end{equation}
 Hence, the negativity gives exactly the same critical value
 of the spin-spin correlation for the absence of entanglement. 
 In fact, any measure of the entanglement shows
 that the critical value of the spin-spin correlation,
 $f^c_s = -1/4$,
 is a unique point for the two impurity Kondo problem. 
 We will see below that the critical correlation can be related to
 the unstable fixed point in the two impurity Kondo model \cite{Jones88}.

\begin{table}
\begin{center}
\begin{tabular}{lll}
 \hline \hline
  & Fidelity $p_s$ &  Correlation $f_s$ \\ \hline
 Concurrence & $p_s \geq 1/2$ &
        $f_s \leq -1/4$ \\
 Quantum teleportation 
 & $p_s \geq 1/2$ \cite{Popescu} & 
        $f_s \leq -1/4$ \\
 Violation of Bell
 -CHSH inequality 
 & $p_s > \left(1+3/{\sqrt{2}}\right)/4 $ \cite{Horodecki95}
 & $f_s < -3/(4\sqrt{2})$ \\
  \hline \hline 
\end{tabular}
\end{center}
\caption{Comparison of criteria for entanglement, quantum teleportation,
         and violation of a Bell inequality in terms of the fidelity, 
         $p_s= 1/4-f_s$, 
         and the spin-spin correlation,
         $f_s \equiv \langle{\bf S}_A\cdot{\bf S}_B\rangle$, 
         for which the two spin $1/2$ 
         is in a spin-rotationally invariant mixed state. 
         Note that the requirement for violation of Bell inequality is
         a more stringent condition then the presence of entanglement.
        }
\label{table}
\end{table}

 \section{Quantum teleportation, Bell inequalities, and correlation}

 There are rigid constraints on the value of the spin-spin correlation
 required to use
 the two Kondo impurities for QIP.
 The state of Eq. (\ref{R3}) for two Kondo impurities is a Werner state that
 is highly symmetric and $SU(2)\otimes SU(2)$-invariant \cite{Werner,Bennett}.
 The Werner state can be entangled 
 but not violate any Bell inequality (i.e., be described by a hidden
 variable theory)
 for some values of the fidelity $p_s$.
 In fact,
 a Werner state with $p_s \leq (1+3/\sqrt{2})/4 \approx 0.78$ satisfies 
 the Clauser-Horne-Shinmony-Holt (CHSH) inequality \cite{CHSH,Horodecki95},
 i.e., does not have the non-local correlations characteristic
 of maximally entangled states.
 This criterion corresponds to 
 $f_s \geq -3/4\sqrt{2} \approx -0.53$  
 in the two impurity Kondo system.
 The values of the spin-spin correlation,
 for an entangled state
 without the violation of the Bell-CHSH inequality,
 is determined by the concurrence/negativity.
 The entangled state, for 
 $ f_s < -1/4$,
 can then be used for QIP 
 including teleportation \cite{Popescu,Lee00}.
 To provide a clear comparison of criteria 
 in terms of the spin-spin correlation
 for the two impurity Kondo problem,
 Table \ref{table} shows values required
 for Bell inequalities, quantum teleportation, and
 entanglement.

\begin{figure}
 \vspace*{9.00cm}
 \includegraphics{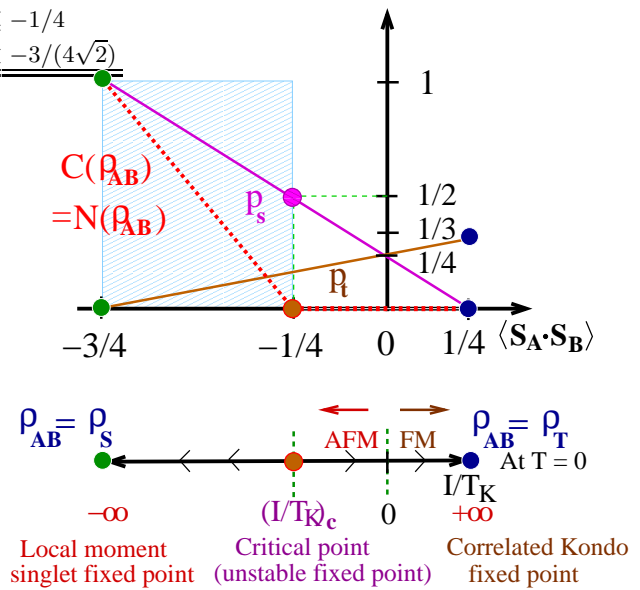}
\caption{(Color online) 
          Top: Relationship between the probabilities for spin singlet $(p_s)$ 
          and triplet $(p_t)$ states between the two Kondo impurity spins 
          ${\bf S}_A$
          and ${\bf S}_B$, and the entanglement measures
          concurrence$(C)$/negativity$(N)$ 
          $(C(\rho_{AB})=N(\rho_{AB})=2p_s-1 \geq 0, \mbox{ otherwise } C=N=0)$ 
          and the spin-spin correlation function
          $f_s \equiv \langle {\bf S}_A \cdot {\bf S}_B\rangle$. 
          Here, $p_s$ corresponds the singlet fidelity.
          Only for 
          $ -3/4 \leq f_s \leq -1/4$, 
          the concurrence/negativity
          has a non-zero value.
          This implies that the critical value of 
          $f^c_s= -1/4$
          separates entangled states (hatched) 
          from disentangled states (non-hatched). 
          The entangled state is useful for quantum teleportation 
          for which the criterion is given by $p_s \geq 1/2$ 
          $(f_s \leq -1/4)$
          \cite{Popescu}.
          Then, for 
          $-3/4\sqrt{2} \leq f_s \leq -1/4$,
          the entangled states which do not violate the Bell-CHSH 
          inequality can be used for quantum teleportation.
          Bottom: Schematic renormalization group flow on the 
          axis of the ratio of RKKY interaction to Kondo temperature, 
          $I/T_K$ at zero temperature taken from Ref. \cite{Jones88}.
          Note that there is a one-to-one correspondence 
          between the fixed point and
          the critical value of the spin-spin correlations.
          }
 \label{summary}
 \end{figure}

 We summarize pictorially the main results of this study;
 the relations between the concurrence/negativity and 
 the probabilities of the states for two impurity spins
 as a function of
 the spin-spin correlation is shown in the top panel of Fig. \ref{summary}.

 \section{Relationship  between concurrence/negativity 
   and quantum phase transitions} 

 A connection between entanglement
 and quantum phase transition(QPT)s has been proposed 
 for a particular class of Hamiltonians \cite{Wu}.
 For first- and second-order QPTs, 
 there occurs a discontinuity in the ground-state concurrence/negativity
 and its first derivative, respectively, due to non-analyticities
 in the ground state energy. 
 In the case of the two impurity Kondo problem, 
 the concurrence/negativity is a continuous function and
 its first derivative
 has a discontinuity at the critical value
 of spin-spin correlation.
 However,
 the discontinuity does not come from non-analyticity 
 in the ground state energy
 but
 from the requirement of non-negative concurrence \cite{Yang}
 or non-positive negativity.
 Consequently, in general, the critical point of the spin-spin correlation
 in the concurrence/negativity
 is not necessarily related to a QPT \cite{Wu}.
 Thus, to use the concurrence/negativity as a signature of QPTs, 
 the non-analyticities in the ground state energy
 should occur 
 at values of spin-spin correlation
 $-3/4 \leq f_s \leq  -1/4$, 
 even if the two impurity Kondo system has a definite QPT.
 Otherwise, 
 since the concurrence/negativity is zero for  
 $-1/4 \leq f_s \leq  1/4$, 
 We see that
 non-analytic behavior of the concurrence/negativity
 is not a definitive signature of a QPT.

 \section{Vanishing entanglement at a quantum critical point}

 We now consider how the vanishing entanglement at
 $f^c_s =- 1/4$ 
 may 
 relate to the unstable fixed point (QPT) of the two-impurity Kondo model 
 found by Jones, Varma, and Wilkins \cite{Jones88}.
 Wilson's numerical renormalization-group technique
 \cite{Jones88} and conformal field theory approaches
 \cite{Ludwig}, have shown that at
 the unstable fixed point, the staggered
 susceptibility and the specific heat coefficient, $\gamma$, diverge,
 this critical value of $(I/T_K)_c$.
 This critical value
 separates the regimes of 
 renormalization group flows to the stable Kondo effect
 fixed point for $ I/T_K > (I/T_K)_c$  and 
 the locked-impurity singlet fixed point for $ I/T_K < (I/T_K)_c$ 
 \cite{Jayaprakash}.
 It should be stressed that this critical point only exists
 when
 there is a symmetry between even and odd parity channels 
 \cite{Jones88,Gan,Silva}.
 The divergence of thermodynamic properties implies that,
 at and around the unstable fixed point,
 a local description of the impurity and conduction electron
 degrees of freedom in terms of a local Fermi-liquid
 is not possible.
 Interestingly, in addition, the spin-spin
 correlation of the ground state varies 
 continuously as a function of $I/T_K$
 and, at the unstable fixed point $(I/T_K)_c$, 
 approaches the critical value of
 $f^c_s = -1/4$
 within numerical accuracy in the wide range of values of $T_K$
 \cite{Jones88}.
 This value at the critical point was also found analytically \cite{Silva}.
 The schematics in the top and bottom panels of Fig. \ref{summary} show
 a correspondence between the entanglement and renormalization
 group flow for the two impurity Kondo system. 
 When the symmetry of even-odd parity is broken 
 the critical point is replaced by a crossover \cite{Gan,Silva}.
 This might suggest that 
 the quantum entanglement for this two impurity Kondo problem
 plays a important role in this quantum phase transition.
 However, if the even-odd symmetry is not present then
 the entanglement still vanishes for a critical value of $(I/T_K)_c$
 but there is no quantum phase transition 
 \cite{Gan,Silva}.

 \section{Entanglement between the conduction electrons 
        and the Kondo impurities} 

 The von Neumann entropy \cite{Nielsen} is a good measure of
 entanglement between two subsystems of a pure state 
 $\left|\Psi_G\right\rangle$ \cite{Bennett96}.
 Although above we considered the total system in terms of the three subsystems
 (two impurity spins $A$ and $B$, and the conduction electrons $C$),
 it can also be regarded as 
 a bipartite system having two subsystems ${\cal A}$ and ${\cal B}$. 
 There are two options
 (i) one impurity spin (${\cal A}=j$) and the remainder of the 
 total system (${\cal B}=j'(\neq j) \cup \; C$) or 
 (ii) two impurity spins (${\cal A}=A\cup B$)
 and the conduction electrons (${\cal B}=C$).
 For the pure state $\left|\Psi_G\right\rangle$
 of the bipartite systems,
 the von Neumann entropy $E(\rho)=-\mathrm{Tr}\; \rho \log \rho$ 
 is given by the density matrix associated with
 either of the two subsystems, i.e., $E(\rho_{\cal A})=E(\rho_{\cal B})$.
 The logarithm is taken in the base 2.

 \begin{figure}
 \vspace*{5.20cm}
 \includegraphics{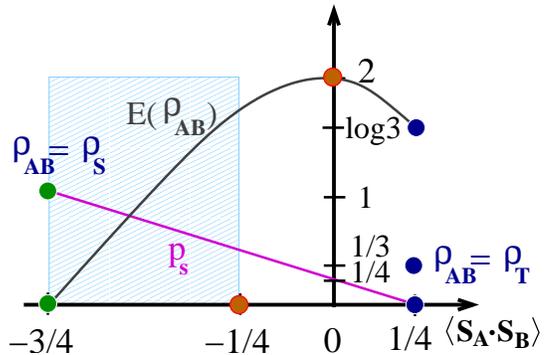}
 \caption{(Color online) 
          The von Neumann entropy $E(\rho_{AB})$ for entanglement between
          the two Kondo spins and the conduction electrons
          is shown
          as a function of the spin-spin correlation function
          $f_s \equiv \langle {\bf S}_A \cdot {\bf S}_B\rangle$. 
          The von Neumann entropy is zero only 
          at $f_s=-3/4$.
          Thus
          the two Kondo spins are entangled with the conduction electrons
          except for the extreme limit of a pure singlet of
          two Kondo spins, i.e., $I/T_K \ll 0$.
          The hatched region for $-3/4 \leq f_s \leq -1/4$
          is compared to show  
          the entanglement between two Kondo spins
          (Compare Fig. \ref{summary}).
          No unique behavior is seen 
          in the von Neumann entropy at $f^c_s=-1/4$.
          When the indirect RKKY interaction disappears at $f_s=0$ $(I=0)$,
          the two Kondo spins are maximally entangled 
          with the conduction electrons.
          At $f_s=1/4$, the von Neumann entropy for
          the triplet states of the two Kondo spins
          is $\log 3$. 
          }
 \label{entropy}
 \end{figure}
 To quantify
 the entanglement of one impurity spin $(j)$ with 
 the remainder $(j'\; C)$ of the total system,
 the reduced (two by two) density matrix of the one impurity spin,
 $\rho_{j}=\mathrm{Tr}_{j'C}\left( \rho \right)  
 =\mathrm{Tr}_{j'}\left( \rho_{AB} \right)$,
 needs to be evaluated
 by taking trace over the state of the remainder of the total system.
 In terms of the Pauli matrices, 
 it has the form  
 $\rho_j = \left(\sigma^0_j + \sum_{\alpha} r_\alpha \sigma^\alpha_j \right)/2$
 with $r_\alpha = \langle \sigma^\alpha_j \rangle$.
 As expected, $\rho_{A} =\rho_{B}=\sigma^0_j/2$ 
 because the expectation value of each impurity spin
 is zero, $\langle \sigma^\alpha_j \rangle=0$,
 due to the spin-rotational invariance of the system.
 Then we have
 \begin{equation}
  E(\rho_j)=-\mathrm{Tr} \rho_j \log_2 \rho_j = 1.
 \end{equation}
 Note that the von Neumann entropy of each impurity spin 
 is not dependent on the spin-spin correlation $f_s$ of the two impurity spins.
 Hence, each Kondo spin is always maximally entangled 
 with the remainder of the total system \cite{Costi}.
 The entanglement of two impurity spins ($A$ and $B$)
 with the conduction electrons ($C$) 
 is quantified by the von Neumann entropy of 
 the reduced density matrix $\rho_{AB}$,
 \begin{equation}
  E(\rho_{AB})=-p_s\log_2 p_s -(1-p_s)\log_2 \frac{1-p_s}{3}. 
 \label{entropy2}
 \end{equation}
 Figure \ref{entropy} shows the von Neumann entropy, $E(\rho_{AB})$,
 and the singlet fidelity, $p_s$,
 as a function of the spin-spin correlation, $f_s$.
 When $f_s=-3/4$ ($p_s=1$),
 $E(\rho_{AB})=0$ and
 the two Kondo spins are completely disentangled from 
 the conduction electrons. 
 The maximum degree of the entanglement of one Kondo spin 
 with the remainder of the total system is then attributed to 
 the other Kondo spin rather than the conduction electron spins.
 The concurrence, $C(\rho_{AB})=1$, as a measure of the entanglement
 between the two Kondo spins shows that 
 they form the AFM spin-singlet state
 in the limit of $I/T_K \ll 0$, as shown in Ref. \cite{Jayaprakash}.
 In the language of Kondo screening, one Kondo spin
 perfectly screens the other Kondo spin
 and the conduction electrons do not participate in screening 
 any Kondo spin.
 As the RKKY interaction increases 
 up to $I/T_K=0$, i.e., $p_s=1/4$,
 the entropy of Eq. (\ref{entropy2}) increases monotonically 
 and reaches its maximum value of 2.
 As discussed, each Kondo spin is always maximally entangled with 
 the conduction electrons but the entanglement of two Kondo spins
 disappears for $f_s \geq -1/4$.
 Thus, partial screenings by 
 one Kondo spin and
 the conduction electrons
 accomplish a complete screening of the other Kondo spin.
 In fact, 
 the competition between
 the Kondo effect, $T_K$, and the RKKY interaction, $I$, determines
 to what extent partial screening 
 of one Kondo spin
 by the other Kondo spin and the conduction electron spins.
 At $f_s=0$ ($p_s=1/4$), i.e., $I/T_K=0$, 
 the two Kondo spins are maximally entangled with the conduction electrons
 but no entanglement between them exists. 
 As the spin-spin correlation (RKKY interaction) 
 increases to $I/T_K \rightarrow \infty$, i.e., $p_s=0$,
 the entropy of Eq. (\ref{entropy2}) decreases gradually 
 to $E(\rho_{AB})=\log_2 3$ in the limit of the FM spin-triplet state.
 As a result,
 the entanglements of (i) one Kondo spin and the remainder of the total system
 and (ii) two Kondo spins and the conduction electrons
 exist in the whole range of the spin-spin correlation,
 and do not show any signatures of the unique behavior 
 of the two impurity Kondo system 
 at the unstable fixed point $(I/T_K)_c$,
 i.e., $f^c_s=-1/4$. 

 \section{Conclusions} 
 Any system of two spins which are a subsystem of a spin-rotationally
 invariant state will have 
 similar entanglement properties.
 To our knowledge, this is the first discussion of 
 an experimental solid-state realization of a Werner state.
 This work has significant implications for proposals using the RKKY
 interaction for QIP.
 We have shown that it is not sufficient to just use the RKKY interaction
 to produce antiferromagnetic correlations between spins.
 Entanglement will only be created when the AFM correlations are larger
 than a critical non-zero value.
 Hence, it is important that realistic estimates be made for the ratio
 $I/T_K$ for candidate systems \cite{Marcus,Piermarocchi,Ardavan}.
 

 \section{Acknowledgments}

  This work was stimulated by discussions with G. A. D. Briggs.
  RHM thanks the QIPIRC at Oxford and Wolfson College for hospitality.
  SYC thanks KIAS for hospitality.
  We thank  John Fjaerestad, John Jefferson,
  Brendon Lovett, Hyunseok Jeong,  
  Gerard Milburn, and Anton Ram\v{s}ak 
  for valuable discussions.
  Andrew Doherty and Yeong-Cherng Liang
  gave us a very helpful introduction to entanglement in Werner states.
  This work was supported by the Australian Research Council.

 
\end{document}